\documentstyle[preprint,aps]{revtex}
\begin{document}
\draft
\title{The Landau theory of phase separation in cuprates}
\author{A.~H.~Castro Neto}
\bigskip
\address
{Institute for Theoretical Physics\\
University of California at Santa Barbara\\
Santa Barbara, CA, 93106-4030}

\maketitle

\begin{abstract}
I discuss the problem of phase separation in cuprates from the point of
view of the Landau theory of Fermi liquids. I calculate the rate of
growth of unstable regions for the hydrodymanics and collisionless limit and,
in presence of long range Coulomb interactions, the size of these regions.
These are analytic results valid for any strength of the Landau parameters.
\end{abstract}

\bigskip


\narrowtext

There is experimental evidence that in the cuprates there exists a phase
separation between hole rich and hole poor regions \cite{emery}.
Also numerical
studies in the $t-J$ model (which is believed to describe the low temperature
physics of these materials) show clearly signs of phase separation \cite{lin}.
It is well established that at zero doping and temperatures higher than the
critical temperature the cuprates can be well described as a Heisenberg
antiferromagnet. This has been confirmed theoretically from the studies of
non-linear sigma models (which reproduces the long wavelength behavior of
the Heisenberg model) and the scaling laws of many measurable quantities
\cite{chakravarty,sokol}. At finite doping holes start to move in the
antiferromagnetic background and interact by exchanging spin waves. At the
level of the $t-J$ model the physics at long wavelengths is described by
the Shraiman-Siggia model \cite{shraiman}. In the last few years Emery and
Kivelson \cite{emery} have also studied the problems of holes in an
antiferromagnetic background and show that the phenomena of phase separation
is quite natural.

In this paper I study the physics of phase separation using the tools of
the Landau theory of Fermi liquids. This work is inspired by the problem of
growth of instabilities in three dimensional Fermi liquids studied by
Pethick and Ravenhall \cite{pethick}. Thus, I will make the assumption
of a Fermi liquid behavior for the holes. This is actually not a strong
assumption because I am going to use a part of the Landau theory which
depends only on conservation laws for the distortions of the Fermi surface.
Recently it was shown that this part of the Landau theory can also be applied
in Luttinger liquids in integrable systems \cite{carmelo}.
In any event, analytic studies of the ferromagnetic region of the
Hubbard model at strong coupling (close to the antiferromagnetic
instability) have shown that the holes form a Fermi liquid \cite{ioffe}.

I will consider a two dimensional system of holes (spinless) with density
$n=\frac{p_F^2}{4 \pi}$ where $p_F$ is the Fermi momentum ($n$ is the
doping density). Since I am interested in the low doping regime I will
assume that the Fermi surface is spherical and I will not take into account
lattice effects. The density of states at the Fermi level is given by
$N(0)=\frac{m^*}{2 \pi}$ where $m^*$ is the effective mass of the
quasi-holes and $m$ is the bare mass.

I will consider angle depend distortions of the Fermi surface (in this
case a line) and parametrize the distortions by an angle $\theta$,
\begin{equation}
\delta n(\theta) = u_0 + 2 \sum_{n=1}^{\infty} u_n \cos(n \theta),
\end{equation}
where I expanded the deviations in Fourier series with coefficients
$u_n$. Also, it is assumed that the quasi-holes interact through the
Landau parameters $f_{\vec{p},\vec{p}'}$ which in principle can be
obtained from the experiment or can be calculated from some theoretical
model (for instance, the $t-J$ model). It is also useful to consider
the dimensionless Landau parameters
$F_{\vec{p},\vec{p}'} = N(0) f_{\vec{p},\vec{p}'}$.

Since the interaction between the particles at the Fermi surface only
depend on the angle between the momenta of the quasi-holes we can also
expand the Landau parameters in Fourier series,
\begin{equation}
F_{\vec{p},\vec{p}'} = F_0 +  2 \sum_{n=1}^{\infty} F_n \cos(n
(\theta-\theta'))
\end{equation}
where $F_n$ are the Fourier components.

At large densities a fermionic liquid is stable against phase separation and
density fluctuations which results in a positive compressibility.
As we decrease the density it can enter a metastable
region where it separates into a liquid-gas phase but it is still
stable against density fluctuations of long wavelengths. In this phase
the compressibility is positive. As the compressibility approaches zero
(the so-called spinodal line) the liquid becomes unstable to long wavelength
density fluctuations as well.
In the context of the Fermi liquid theory this
result can be understood from the expression for the sound velocity, $c_S$,
in a Fermi liquid which is given by \cite{baym},
\begin{equation}
c_S^2= \frac{n (1+F_0)}{m N(0)} = \frac{m^*}{m} \frac{1+F_0}{2} v_F^2,
\end{equation}
where $v_F =\frac{p_F}{m^*}$ is the Fermi velocity.
Observe that if $F_0 < -1$ then $c_S^2 < 0$ and an unstable collective
mode grows in the system and there is a phase separation. For the moment
I am considering as a starting point a homogeneous system and local
interactions. It means that the instability grows in the system as a
whole. If we include the Coulomb interaction, as I will show later, then
the instability has a finite size. For the moment we consider local
interactions since the inclusion of Coulomb interactions is trivial.

It is also well established that in the cuprates the effective mass
of the holes can be many times the bare mass. In the context
of the Fermi liquid theory this is due to the a non-isotropic interaction
on the Fermi surface. Indeed, the relation between the effective mass
and the bare mass is given by \cite{baym},
\begin{eqnarray}
\frac{m^*}{m} = 1 - \sum_{\vec{p}'} f_{\vec{p},\vec{p}'} \frac{\partial
n^0(p')}{\partial \epsilon(p')} \frac{\vec{p} \cdot \vec{p}'}{p^2}
\nonumber
\end{eqnarray}
where $n^0(p)=\Theta(\mu - \epsilon(p))$ is the zero temperature distribution
for the holes and $\mu$ is the chemical potential. Substituting (2) and
using the orthogonality relation between the Fourier components one finds,
\begin{equation}
\frac{m^*}{m}  = 1 + F_1 .
\end{equation}
Therefore, the change in the effective mass depends only on the second
Fourier component. From now on we will keep only $F_0$ (related to the
instability) and $F_1$ (related to the increase in the mass) since these are
the main ingredients for the physics of the system.

Observe that for $F_0 < -1$ the hydrodynamical rate of growth of the
instability in the system (which determines the time scale for the growth,
$\Gamma^{-1}$) is obtained straight from (3),
\begin{equation}
\Gamma(q) = \left(\frac{(1+F_1) |1+F_0|}{2}\right)^{1/2} q v_F.
\end{equation}
This result is valid when the collisions dominate the physics, that is,
the mean free path is short. However, since the density of holes is
small the mean free path can be quite large and the collisionless limit
is more interesting.

The collisionless limit can be obtained from the Landau equation for
sound waves in the absence of a collision integral (which can also be
included \cite{next}). It reads \cite{baym},
\begin{equation}
\frac{\partial \delta n_{\vec{p}}}{\partial t} + \vec{v}_{\vec{p}}
\cdot \nabla  \delta n_{\vec{p}} - \frac{\partial n^0(p)}{\partial \epsilon(p)}
\vec{v}_{\vec{p}} \cdot  \nabla \left( U + \sum_{\vec{p}'} f_{\vec{p},\vec{p}'}
 \delta n_{\vec{p}}\right) = 0,
\end{equation}
where $U$ is an external field and in terms of the deviations (1) we have
$ \delta n_{\vec{p}} = \frac{\delta n(\theta)}{N(0)} \left( - \frac{\partial
n^0(p)}{\partial \epsilon(p)}\right)$.

Assuming an harmonic evolution in time, that is,
$\delta n(\theta) \sim e^{i(\vec{q} \cdot \vec{r}- \omega t)}$,
the Landau equation has the form,
\begin{equation}
\left(q v_F \cos(\theta)-\omega \right) \delta n(\theta) + q v_F  \cos(\theta)
\left(N(0) U + \int_{0}^{2 \pi} \frac{d \theta'}{2 \pi} F(\theta-\theta')
\delta n(\theta')\right) = 0,
\end{equation}
 which can be simplified by our assumption of keeping only the two first
Fourier components,
\begin{equation}
\left(q v_F \cos(\theta)-\omega \right) \delta n(\theta) + q v_F  \cos(\theta)
\
\left(N(0) U + F_0 u_0 + 2 F_1 u_1  \cos(\theta)\right) = 0.
\end{equation}

Integrating (8) over $\theta$ one finds,
\begin{equation}
u_1 = \frac{\omega u_0}{q v_F (1+F_1)},
\end{equation}
which is essentially the conservation of the number of particles.

Solving (8) for $\delta n(\theta)$ one gets,
\begin{equation}
\delta n(\theta) = \frac{q v_F \cos(\theta)}{\omega-q v_F \cos(\theta)} \left(
N(0) U + F_0 + 2 F_1 u_1  \cos(\theta)\right),
\end{equation}
and an integration over $\theta$ leads to,
\begin{equation}
u_0 = \chi_{RPA}(s) \left(N(0) U + F_0 u_0 + 2 s F_1 u_1\right)
\end{equation}
where $s = \frac{\omega}{q v_F}$ and
\begin{equation}
 \chi_{RPA}(s) = \int_{0}^{2 \pi} \frac{d \theta}{2 \pi}
\frac{\cos(\theta)}{s-\cos(\theta)} = \frac{s}{\sqrt{s^2-1}}-1
\end{equation}
is the RPA density-density correlation function \cite{baym}.

Since we are interested in the growth of collective modes we calculate the
density-density correlation function for the problem,
\begin{equation}
\chi(s)= \frac{u_0}{N(0) U} =
\frac{ \chi_{RPA}(s)}{1- \chi_{RPA}(s) \left(F_0 + s^2 A_1\right)}
\end{equation}
where,
\begin{equation}
A_1 = \frac{2 F_1}{1+F_1}.
\end{equation}

As it is well known the collective modes are given by the poles of the
correlation function (13), that is, the points where $\chi^{-1}=0$.
Unlike the three dimensional case \cite{pethick},
which leads to a transcendental equation
that can be solved only numerically, in two dimensions we can solve
this problem analytically for any strength of the interaction. The equation
has two roots which read,
\begin{equation}
s_{\pm}^2 = \frac{-1+2 A_1-2 F_0+2 A_1 F_0 \pm
\sqrt{1+4 A_1+4 F_0+4 F_0^2 + 4 A_1 F_0}}{2 A_1 (2-A_1)}.
\end{equation}
It is easy to conclude that $s_+^2 \geq 0$ and therefore is related to
a true collective mode in the system and $s_-^2 \leq 0$ which means
that the rate of growth of the instability in given by,
\begin{equation}
\Gamma(q) = s_- q v_F.
\end{equation}

The inclusion of Coulomb interaction, $V(r) = \frac{e^2}{ \epsilon r}$
(where $\epsilon$ is the dielectric constant), is
rather trivial since the interaction is isotropic. The Fourier transform
of this interactions in two dimensions is given by
$V(q) = \frac{2 \pi e^2}{\epsilon q}$
and following the prescription of the Landau
theory \cite{baym} the only change in our calculation would be to make
the substitution, $F_0 \to F_0 + \frac{2 \pi e^2 N(0)}{\epsilon q} $.
Observe
that this modification would lead to a momentum depend relation for
the build up of the instability ($F_0 + \frac{2 \pi e^2 N(0)}{\epsilon q}
<-1$) namely,
\begin{equation}
q \geq \frac{e^2 m (1+F_1)}{\epsilon |1+F_0|}.
\end{equation}
This relation means that the instability can be probed only for
wavelengths smaller than $\frac{\epsilon |1+F_0|}{e^2 m (1+F_1)}$ which
established the minimum size for the unstable regions in the plane,
that is, the size of the role rich regions.
Of course, in the absence of Coulomb forces ($e=0$) the instability
region has the size of the system.

In conclusion, in this paper I use the Landau theory of Fermi liquids
in order to calculate analytically the rate of growth of phase separation
in a two dimensional systems of holes in cuprates for any strength of the
Landau parameters. I also estimate the size of the unstable regions due
to the Coulomb repulsion between the holes.

I would like to acknowledge S.A.~Kivelson, H.Q.~Lin,
E.~Miranda and H.~Monien for many illuminating comments.
This research was supported in part by the National Science Foundation
under the Grant No. PHY89-04035.

\end{document}